\documentclass[conference]{IEEEtran}
\usepackage[letterpaper, left=1in, right=1in, bottom=1in, top=0.75in]{geometry}
\usepackage{booktabs} 
\usepackage{algorithm}
\usepackage[noend]{algpseudocode}
\usepackage[inline]{enumitem}
\usepackage{multirow}
\usepackage{color}
\usepackage{graphicx}

\usepackage[cmex10]{amsmath}
\usepackage{svg}
\usepackage{amssymb}
\usepackage{amsthm}
\usepackage{amsmath}
\usepackage{mathtools}
\usepackage{float}
\usepackage{tikz}
\usepackage{epstopdf}
\usepackage{comment}

\graphicspath{ {figures/} }

\DeclareMathOperator*{\argmax}{arg\,max}
\DeclareMathOperator*{\argmin}{arg\,min}

\newcommand{\saeed}[1]{}

\newcommand{\wooramdel}[1]{}

\newcommand{\doc}[1]{}

\newcommand{\smallcell}{SmallC} 
\newcommand{\smallcells}{SmallCs} 

\newcommand{\macrocell}{MacroC} 
\newcommand{\macrocells}{MacroCs} 

\newcommand{\servcell}{ServeC} 
\newcommand{\servcells}{ServeCs} 

\newcommand{\primecell}{PrimeC} 

\newcommand{\secondcell}{SecondC} 

\usetikzlibrary{chains,shapes.multipart}
\usetikzlibrary{shapes,calc,fit}
\usetikzlibrary{automata,positioning}
\tikzset{
queue/.pic={
  \draw[line width=0.7pt] (0,0) -- ++(1cm,0) -- ++(0,-0.7cm) -- ++(-1cm,0);
    \foreach \i in {1,...,4}
      \draw[line width=0.5pt] (1cm-\i*4pt,0) -- +(0,-0.7cm);
    \draw[line width=0.7pt] (1.35,-0.35cm) circle [radius=0.35cm];
    \node[align=center] at (1.35,-0.35cm) {$\mu$};
    \node[above] at (0.7cm,0) {$#1$};
  }
}

\title{Predictive Network Control in Multi-Connectivity Mobility for URLLC Services\\ }
\date{Started: 11.03.2019, last edit: \today}
\author{ \\ \IEEEauthorblockN{David Guzman, Richard Schoeffauer, and Gerhard Wunder}
\IEEEauthorblockA{Freie Universit{\"a}t Berlin, Germany\\
\{david.guzman, richard.schoeffauer, g.wunder\}@fu-berlin.de}}

\begin{document}
\maketitle

\begin{abstract}
This paper proposes a centralized predictive flow controller to handle multi-connectivity for ultra-reliable low latency communication (URLLC) services.
The prediction is based on channel state information (CSI) and buffer state reports from the system nodes. For this, we extend CSI availability to a packet data convergence protocol (PDCP) controller. The controller captures CSI in a discrete time Markov chain (DTMC). The DTMC is used to predict forwarding decisions over a finite time horizon. The novel mathematical model optimizes over finite trajectories based on a linear program.
The results show performance improvements in a multi-layer small cell mobility scenario in terms of end-to-end (E2E) throughput. Furthermore, 5G new radio (NR) complaint system level simulations (SLS) and results are shown for dual connectivity as well as for the general multi-connectivity case.
\end{abstract}

\section{Introduction}
\label{sec:intro}
Enhanced mobile broadband (eMBB), URLLC and massive machine type communication (mMTC) are services introduced by the recently released 5G NR, 3GPP Release 15. One of the most challenging goals for an URLLC service is to target 99.999\% reliability at a millisecond level latency \cite{TS38300}. These performance requirements for URLLC services are critical when multi-connectivity (MC) mobility in heterogeneous networks (HetNets) is considered.

HetNets rely on increasing the number of small cells (\smallcells{}) to supplement already installed macro cells (\macrocells{}), and bringing them closer to the user equipment (UE) resulting in improvements of capacity per area \cite{Pedersen001}. HetNets, comprising \macrocells{} and \smallcells{}, with MC offer the possibility to use several variants of multi-cell cooperation and coordination techniques. Hence, extensive studies have been carried out for HetNets with MC in terms of architecture, mobility, and connectivity, aiming to increase reliability, reduce latency, and diminish mobility event durations and interruption times.

Architectural optimizations for handover procedures and \macrocell{} mobility based on the synchronicity of the cells and radio access network (RAN) layers virtualization are presented in \cite{Kolding} and \cite{Barbera002}. Furthermore, in \cite{Barbera001}, a mobility performance sensitivity analysis is presented for dual connectivity (DC) HetNets, giving trends and observations from simulations. Mobility enhancements in multi-layer networks are revised in \cite{Pedersen001} where a hybrid mobility solution is proposed. The hybrid mobility concept relies on a network-controlled \macrocell{} mobility and an UE-autonomous \smallcell{} mobility.

Moreover, DC, as an initial case of MC, is studied in \cite{Rosa001}, where its architecture, functionalities and performance aspects are presented. MC, as a novel feature to enhance reliability and reduce the packet failure probability (by transmitting data packets to the UE from multiple cells independently), is considered in \cite{Mahmood}. In addition, an admission control mechanism based on mainly the user's channel quality is designed. Although the proposed mechanism is highly sensitive to the configuration criteria \cite{Mahmood}, latency reduction and reliability improvements compared to single connectivity are observed. Furthermore, in the context of DC HetNet connectivity, a flow control algorithm for the connections between \macrocells{} and \smallcells{} is shown in \cite{Wang}. The flow control mechanism is based on minimizing data buffering time in the \smallcells{} to exploit DC.
\par Our paper focuses on URLLC service provision in an MC mobility scenario where \macrocells{} and \smallcells{} are deployed at non-overlapping carrier frequencies. We consider \smallcell{} mobility events in an established multi-connection, as well as a split bearer architecture to deal with the user plane data flow. Split bearer architecture is a promising enabler to cope with cell densification and MC complexity, which leads us to centralize the control of the system in a single entity. 
\par Towards this end, a centralized predictive controller is
proposed consisting of optimized discrete time binary actions and a policy governed by the wireless channel evolution. The latter is captured in a DTMC based on CSI. And, the control actions are computed by solving a linear program over a finite time horizon. The linear program formulation significantly reduces complexity as well as relaxes the performance sensitivity on the system evolution. The analytical bounds given by physical layer events are shown for the algorithm, which indicates that in a multi-layer mobility setting for URLLC restrictions the proposed algorithm outperforms previous approaches. The performance of the solution is evaluated through 3GPP release 15 compliant SLS, and supported by extensive analysis of lower layer performance indicators. 


\textbf{Organization.} The remainder of the paper is organized as follows: First, a MC, mobility and CQI feedback generation overview is outlined in Section \ref{sec:overview} to further introduce the prediction model in Section \ref{sec:model}. The implementation of the policy is presented in Section \ref{sec:contribution}. Section \ref{sec:simulation} explains the adopted methodology to perform SLS, and Section \ref{sec:performance} shows the performance results. Finally, we draw conclusions and outline further work in Section \ref{sec:conclusion}.

\textbf{Notation.}  Vectors and matrices are written as follows: $\mathbf{a} = (a_1,...a_n), \mathbf{A} = (a_{ij})$ where $\operatorname{diag}(a_{ij})$ stands for a diagonal matrix, and $\mathbf{a}^\intercal$ is the transpose of $\mathbf{a}$. Moreover, $\lceil . \rceil$ is the ceil function and $|\mathcal{R}|$ is the cardinality of a countable set $\mathcal{R}$.

\section{ Multi-connectivity, Small Cell Mobility, and CQI Feedback Generation}
\label{sec:overview}

\subsection{Multi-connectivity and Small Cell Mobility}
\label{subsec:smobility}
\par
\begin{figure}[th]
	\centering
	\includegraphics[width=\columnwidth]{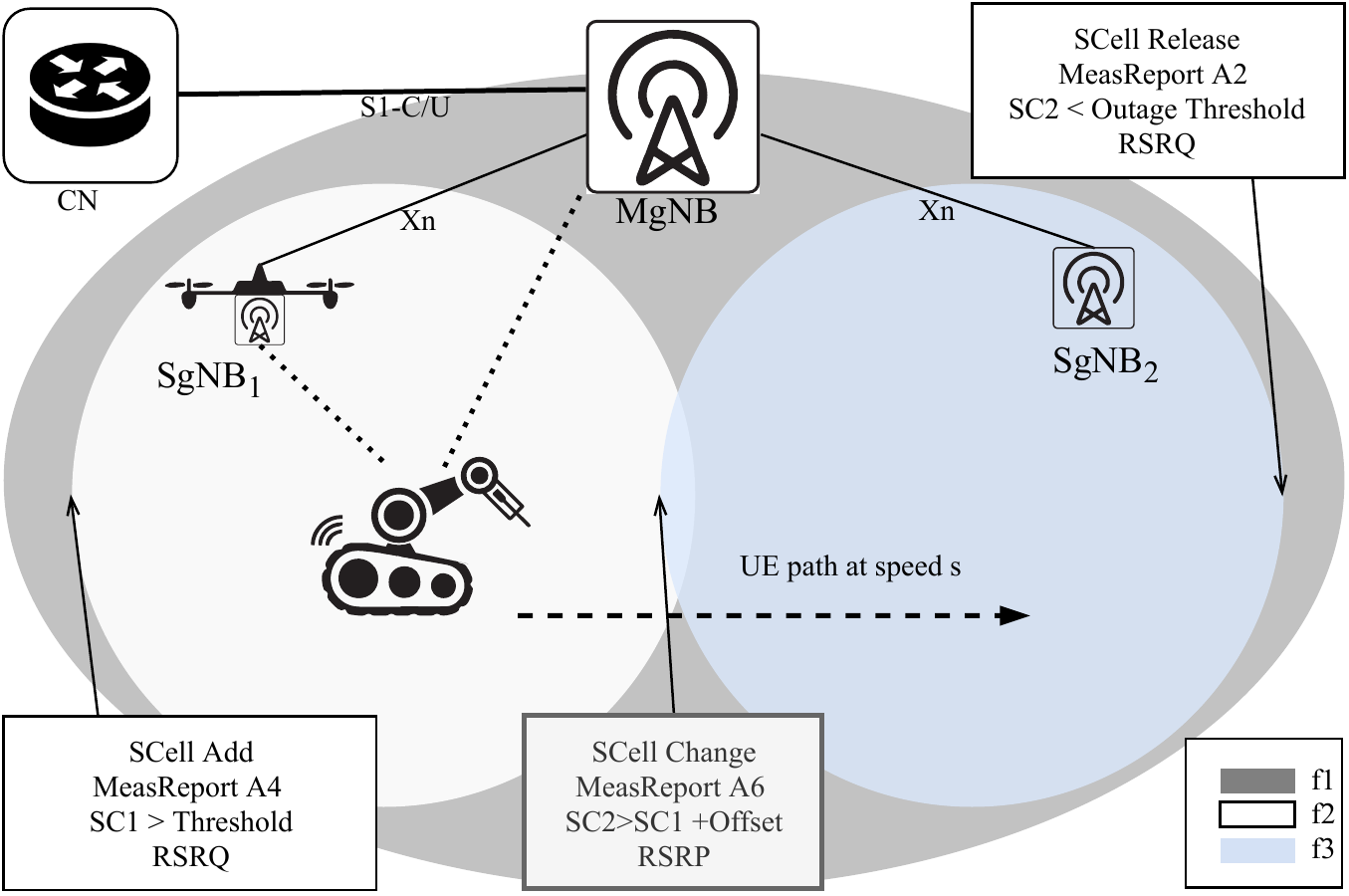}
	\caption{Small Cell Mobility Events and MC in a HetNet}
	\label{fig:small-cell-mobility}
\end{figure}

 In a scenario where \macrocells{} and \smallcells{} are deployed at non-overlapping carrier frequencies $f_1<f_2<f_3$, as shown in Fig. \ref{fig:small-cell-mobility}, UEs can configure MC in terms of inter-site carrier aggregation. Hence, the UE can be served in a \macrocell{} by a master NR gNB (MgNB) and in an \smallcell{} by a secondary NR gNB (SgNB) simultaneously. The \smallcells{} are placed along urban streets, and we use a frequency reuse scheme deployment as proposed in \cite[Scheme 4]{Polignano} in order to avoid inter-cell interference. The MgNB establishes a control interface to the core network (CN), while both the MgNB and the SgNB manage the radio resource control (RRC) in the UE \cite{Barbera001}. 
\par Moreover, the MgNB RRC connection is established before MC, and is used to configure and control CSI feedback and channel measurements in the UE. The reports from the UE include detected cells and CSI to facilitate rate adaptation, cell selection and MC configuration. Based on the channel measurements, the MgNB can set up MC by requesting the SgNB to allocate resources for a specific UE \cite{Mahmood}. The SgNB data connection is restricted to the instantaneous available physical resources. A control plane can be established from any node (MgNB and/or SgNB) to the UE; however, the data plane interface is configured to either MgNB or SgNB but not both, in accordance with the so-called split bearer architecture \cite{Rosa001}. The control and data plane from the CN to the MgNB use the S1 interface as indicated in Fig. \ref{fig:small-cell-mobility}.

The UE's assistive behaviour is based on measurements of the reference signal received power (RSRP) or reference signal received quality (RSRQ) from \macrocells{} and \smallcells{} \cite{Barbera001}. Both are based on the received signal strength indicator (RSSI), which is the total received wide-band power, measured by the UE. Following \cite{TS36133}, the RSSI of the $k$-th UE can be written as
\begin{equation}
	\mathsf{RSSI}_{k} = P_{\mathsf{\servcell{}}} + P_{\mathsf{cc}} + P_{\mathsf{no}}
\end{equation}
where $P_{\mathsf{\servcell{}}}, P_{\mathsf{cc}}, P_{\mathsf{no}}$ are serving cell (\servcell{}), neighbour co-channel cells, and thermal noise powers, respectively. Let $\mathcal{P}$ be the set of physical resource blocks (PRBs) per channel bandwidth. Then the RSRP is a linear average of a reference signal (RS) power, measured from a cell \cite{TS36133}, and is computed as:
\begin{equation}
	\mathsf{RSRP}_{k}=\mathsf{RSSI}_{k} - 10\log(12\cdot|\mathcal{P}|)
\end{equation}

The user $k$ feedbacks an RSRP report as $\mathsf{RSRP}_k = \mathsf{MEAS}^{\mathsf{RSRP}}_i$, if $\mathsf{RSRP}_{k} \in [\mathsf{RP}_{i-1}, \mathsf{RP}_{i})$, where $\mathsf{RP}_0, \mathsf{RP}_1,...,\mathsf{RP}_{n-1}, \mathsf{RP}_{n}$ are the $n$-th RSRP thresholds in dBm \cite{TS36133}. Here, $\mathsf{MEAS}^{\mathsf{RSRP}}_i$ is an input for the L3 measurement layer for cell quality as in \cite[Sec. 9.2.4]{TS38300}. L3 aims to filter the measurements before evaluation as follows \cite[Sec. 5.5.3.2]{TS38331}: 
\begin{equation}
	F^{P}_{t}= \left(1-\left(\frac{1}{2}\right)^{\frac{p}{4}} \right)F^{P}_{t-1}\left(\frac{1}{2}\right)^{\frac{p}{4}}\mathsf{MEAS}^{\mathsf{RSRP}}_{i,t}
\end{equation}   
where $F^{P}_{t}$ is the updated filtered result, $F^{P}_{t-1}$ is the previous measurement result, $p$ is a filter coefficient configured by the RRC signaling, and $\mathsf{MEAS}^{\mathsf{RSRP}}_{i,t}$ is the latest measurement received from the PHY layer.

The RSRQ is a measure of signal power and interference, and for a UE $k$ in a cell $c$ is calculated as:
\begin{equation}
	\mathsf{RSRQ}_{c,k}=\frac{|\mathcal{P}|\cdot \mathsf{RSRP}_{k}}{\mathsf{RSSI}_{k}}
\end{equation}   
A UE sends an RSRQ report to the \servcell{} as $\mathsf{RSRQ}_{c,k}=\mathsf{RSRQ}_i,$ if $\mathsf{RSRQ}_{c,k} \in [\mathsf{RQ}_{i-1},\mathsf{RQ}_{i})$ where $\mathsf{RQ}_0,\mathsf{RQ}_1,...,\mathsf{RQ}_{n-1},\mathsf{RQ}_{n}$ are the $n$-th thresholds in dB.

Let us come back to the scenario depicted in Fig. \ref{fig:small-cell-mobility}, with a set $\mathcal{M}$ of \macrocells{}, and a set $\mathcal{S}$ of \smallcells{}. The \servcells{} for the UE are determined, based on UE measurements of the received downlink RSRQ \cite{Wang}. The primary cell (\primecell{}) $c_k$ for the UE $k$ is selected as: 
\begin{equation}
	c_{k}= \argmax_{c \in \mathcal{M} \cup \mathcal{S}}(\mathsf{RSRQ}_{c,k}+ \mathsf{RE}^{c})
\end{equation}
where $\mathcal{M \cup S}$ is the set of candidate cells and $\mathsf{RE}^{c}$ is the range extension (RE) offset applied to cell $c$. $\mathsf{RE}^{c} = 0$ for $c \in \mathcal{M}$ and $\mathsf{RE}^{c} \geq 0$ for $c \in \mathcal{S}$.

A UE, configured with DC/MC as shown in Fig. \ref{fig:small-cell-mobility}, has a baseline \primecell{} and, when feasible, requests a secondary cell (\secondcell) through an RSRQ A4 event. In an RSRQ A4 event, a neighbouring \smallcell{} becomes better than a threshold (\smallcell{} Add); in an RSRQ A2 event, a \servcell{} becomes worse than a threshold (\smallcell{} Release). Hence, a \servcell{} $e_k$ for a UE $k$ is selected as: 
\begin{equation}
\label{eq:scellselection}
	e_k =
	\begin{cases}
	 \argmax\limits_{c \in \mathcal{S}}(\mathsf{RSRQ}_{c,k}) & $if $ \exists c \in \mathcal{S} : \mathsf{RSRQ}_{c,k} > \mathsf{RQ}^{\mathsf{\servcell{}}} \\
	 \phi & $otherwise$
	\end{cases}
\end{equation}
where $\mathsf{RQ}^{\mathsf{\servcell{}}}$ is a fixed RSRQ \servcell{} threshold. 

When DC is the baseline and a neighbouring \smallcell{} becomes offset better than the current \servcell{}, the UE considers RSRP $A6^{1}$ and $A6^{2}$ conditions. $A6^{1}_t$ is an entering condition, and $A6^{2}_t$ a leaving condition expressed as follows \cite{TS38331}:
\begin{equation}
\label{eq:a61}
	A6^{1}_t : F^{n,P}_{t} + O^{n,P} - Y^{n,P} > F^{s,P}_{t} + O^{s,P} - A6^{\mathsf{off}} 
\end{equation}
\begin{equation}
\label{eq:a62}
	A6^{2}_t : F^{n,P}_{t} + O^{n,P} + Y^{n,P} < F^{s,P}_{t} + O^{s,P} + A6^{\mathsf{off}}
\end{equation}
where $F^{n,P}_{t}$ and $F^{s,P}_{t}$ are the filtered RSRP in dBm for the neighbouring \smallcell{} and serving \smallcell{} respectively. The serving \smallcell{} is selected as in Eq. \ref{eq:scellselection}. Furthermore, $O^{n,P}$ and $O^{s,P}$ are the offsets for neighbouring and serving \smallcells{}, $Y^{n,P}$ is the hysteresis for A6 event, and $A6^{\mathbf{off}}$ is an event offset \cite{TS38331}. All offsets and hysteresis are expressed in dB.

\subsection{CQI Feedback Generation}
\label{subsec:cqi}

The channel quality indicator (CQI), a type of CSI in the RAN, is a 4-bit value that indexes an estimated effective SNR \cite{Donthi_EESM} \cite{TS36213}. The SNR measurements are performed over e.g. CSI-RS control signals broadcasted by the gNBs. The CQI indices and their interpretations are given in \cite[Tbl. 7.2.3-1, Tbl. 7.2.3-5, Tbl. 7.2.3-6]{TS36213} for reporting CQI based on QPSK, 16QAM and 64QAM. 
The gNB controls the CQI reports according to the upper layer transmission mode for \macrocells{} and \smallcells{} \cite{TS36213}. The maximum resolution of CQI in terms of frequency is a sub-band which consists of $p$ contiguous PRBs, where $2 \geq p \geq 8$. The total number of sub-bands is $S= \lceil |\mathcal{P}|/p \rceil$, and the $\mathcal{P}$ in a sub-band $s$ is $\mathcal{P}(s)$. 
    
The report schemes can be either \textit{UE selected sub-band CQI feedback}, or \textit{sub-band-level CQI feedback}. Both schemes use effective exponential SNR mapping (EESM) to generate the CQI value(s) of the PRBs that constitute the sub-bands \cite{Donthi_EESM}. EESM has been extensively studied in \cite{Donthi_EESM}, \cite{Hanzaz01} and \cite{Hanzaz02}. In a UE selected sub-band feedback scheme, the sub-band SNR, $\gamma^{\text{sub-band}}_{s,k}$ of the $k$-th UE for sub-band $s$ is the effective SNR over its constituent $\mathcal{P}$:   
\begin{equation}
	\gamma^{\text{sub-band}}_{s,k}= -\theta \ln\left(\frac{1}{p} \sum_{n \in \mathcal{P}(s)}\exp\left(-\frac{\gamma_{n,k}}{\theta}\right)\right)
\end{equation}
where $\theta$ is a tuning parameter function of upper layer protocol data unit (PDU) length and modulation and coding scheme (MCS).

The $k$-th UE, in connected mode, sorts the sub-band SNRs of its $S^{\mathsf{\smallcell{}}}$ and $S^{\mathsf{\macrocell{}}}$ sub-bands for \smallcell{} and \macrocell{} respectively; for example, sub-bands in a \macrocell{} as $\gamma^{\text{sub-band}}_{(1),k}  \geq ... \geq\ \gamma^{\text{sub-band}}_{(M),k} \geq ... \geq\ \gamma^{\text{sub-band}}_{(S^{\mathsf{\macrocell{}}}),k}$.Then, the UE reports a subset $\mathcal{I}_k = \{1,...,M\}$ which contains the indexes of the $M$ sub-bands with the highest CQIs, as well as a single CQI $R^{\text{highest}}_k$. Indeed, $R^{\text{highest}}_k$ is a function of EESM over the $M$ selected sub-bands \cite{Donthi_EESM}:
\begin{equation}
\label{eq:effectivesinr}
	\gamma^{\text{effective}}_{k}= -\theta \ln\left(\frac{1}{M} \displaystyle\sum_{r \in \mathcal{R}_k}\exp\left(-\frac{\gamma^{\text{sub-band}}_{r,k}}{\theta}\right)\right)
\end{equation} 

If ${\gamma^{\text{effective}}_{k} \in [ \mathsf{LR}_{i-1}, \mathsf{LR}_i)}$, the reported value is $\mathcal{R}^{\text{highest}}_k = r_i$, where $\mathsf{LR}_{i}$ are the link adaptation thresholds, assuring a pre-determined block error rate (BLER) at the chosen MCS index \cite{TS36213}.
At the gNB, $\mathcal{R}^{\text{highest}}_k$ is mapped to an MCS index based on the downlink control indicator (DCI) format. The MCS index determines transport block sizes (TBS) and code block sizes (CBS) to achieve a certain BLER. The TBS depends on the number of assigned PRBs, $|\mathcal{P}|$ \cite{TS36213}. 

%
\section{Prediction Model}
\label{sec:model}
In this paper, in order to study forwarding of PDCP PDUs, we consider a 5G-NR downlink system where there are $\mathcal{M} \cup \mathcal{S}$ cells (\macrocells{} and \smallcells{}), and $K$ uniformly distributed UEs per cell. Users are multiplexed by the orthogonal frequency division multiple access (OFDMA). Moreover, we consider a URLLC traffic model with finite payload of $B$ bytes and Poisson arrival process. URLLC traffic is scheduled with a short TTI of 2-OFDM symbols due to its latency constraints. The rest of non-URLLC traffic is scheduled with variable TTI, the longest consisting of 14-OFDM symbols. Due to the maximum resolution given by the sub-band CQI, we consider a sub-band $S$ as the smallest scheduling unit in the frequency domain. The sub-band is the collection of the PRBs of 12 sub-carriers with 15 kHz spacing corresponding to numerology $0$.
\begin{figure}[th]
	\centering
    \begin{tikzpicture}[>=latex,font=\footnotesize]
    \path 
      (0,0cm) pic {queue=\mathsf{MgNB}\; q_0}
      (2.7,2cm) pic {queue=\mathsf{SgNB}_2\; q_2}
      (2.7,-2cm) pic {queue=\mathsf{SgNB}_1\; q_1}
      (5.4,0cm) pic {queue=\mathsf{UE}\; q_3};
    \draw[<-] (0,-0.35) -- +(-10pt,0) node[left] {$\lambda$};
    \draw[->] (1.7,-0.35) -- +(1,2cm) node[midway,right]{$l_3$} node[midway,left]{$\mathsf{Xn}$};
    \node[align=center] at (2,0.95) {};
    \draw[->] (1.7,-0.35) -- +(1,-2cm)node[midway,right]{$l_1$}node[midway,left]{$\mathsf{Xn}$};
    \draw[dashed,->] (1.7,-0.35cm) -- (5.5,-0.35cm) node[midway,above]{$l_0$};
    \draw[dashed,->] (4.4,1.65cm) -- (5.5,-0.35cm) node[midway,right]{$l_4$};
    \draw[dashed,->] (4.4,-2.35cm) -- (5.5,-0.35cm) node[midway,right]{$l_2$};
    \node[draw,circle,double,double distance=2.5pt,inner sep=0.2cm](control)at (0.5,-2cm){PNC} ;
    \draw[->] (0.5,-1.4) -- +(0,0.7cm);
    \end{tikzpicture}
    \caption{Prediction Model}
	\label{fig:system}
\end{figure}
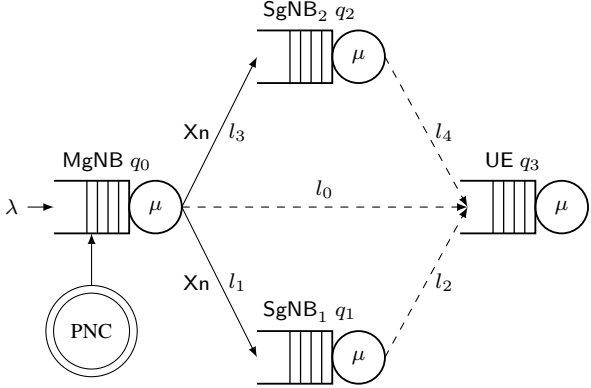

We focus on a routing problem modeled by a discrete time, transport block level queuing system as depicted in Fig. \ref{fig:system}. A single time step is considered to correspond to a TTI of 2-OFDM symbols. The model represents the situation that occurs right after a \smallcell{} Change event, which in turn is based on a $A6^{1}_t$ report. Each network agent is represented by a queue, that holds data packets which are intended for the UE. The queues are connected with wired and wireless links (solid and dashed arrows respectively).
The system state at time $t$ can be represented by the queue vector $\mathbf{q}_t = (q_{0,t},...,q_{3,t})^\intercal \in \mathbb{N}^4$ that contains all $4$ queues. A corresponding arrival vector $\mathbf{a}_t = (a_{0,t},...,a_{3,t})^\intercal \in \mathbb{N}^4$ of stochastic nature models external arrival of data to the system, which in our case only concerns the MgNB buffer queue $q_{0,t}$.

The $5$ links $\mathbf{l}_{0} ,\dots \mathbf{l}_{4}$ are represented by column vectors in such a way that e.g. $\mathbf{l}_{0,t} = (-T, T, 0,...,0)^\intercal$ connects $q_0$ to $q_1$. Here, $T$ is a TBS determined by look up tables (LUTs) by the MCS index, $\mathcal{I}_{MCS}$, and $|\mathcal{P}|$  \cite[Tbl. 7.1.7.2.1-*]{TS36213}. All links are collected in the routing matrix $\mathbf{R} \in \mathbb{Z}^{4 \times 5}$.

A link in $\mathbf{R}$ can be activated by a binary control vector $\mathbf{u}_t \in \{0,1\}^5$, to initiate a transmission along its path. However, due to short term wireless characteristics like channel fading, such a transmission only succeeds with probability $m_{i,t}$, $i = 0,\dots 4$. If we collect these probabilities in a diagonal matrix $\mathbf{M}_t = \operatorname{diag} \{ m_{0,t},\dots m_{4,t} \} $, the system evolution can be written as \cite{PNC}:      
\begin{equation}
\label{eq:model}
	\mathbf{q}_{t+1} = \mathbf{q}_{t} 
	+ \mathbf{R} \mathbb{B} \left[ \mathbf{M}_t \right] \mathbf{u}_t 
	+ \mathbf{a}_t
\end{equation}
where a Bernoulli trial $\mathbb{B}$ on $\mathbf{M}_t$ models success and failure of initiated transmissions.

The control vector is mainly restricted by the constituency matrix (interference constraint) $\mathbf{A} = \begin{bmatrix} 0 & 0 & 1 & 1 & 1 \end{bmatrix}$ and the positiveness of the queues \cite{PNC} \cite{Schoeffauer002}, leading to
\begin{equation}
\label{eq:model_r1}
	\mathbf{u}_{t}\in\mathcal{U}_t:= \left\{
	\mathbf{v}_{t} \in \{0,1\}^{5} \middle| 
	\begin{array}{c}
	\mathbf{A}\mathbf{v}_t \leq \mathbf{1}
	\\
	\mathbf{q}_{t} + \mathbf{R} \mathbf{v}_t \geq \mathbf{0}
	\end{array}
	\right\}
\end{equation} 

As indicated, the evolution of the wireless channels is captured in time dependent success probabilities that are the diagonal elements of $\mathbf{M}_{t}$.
As the main premise for the prediction, we assume existence and knowledge of a DTMC, that governs the evolution of $M_t$ over time. For this purpose, we set up a finite set of probability matrices $\mathbf{M}^1 , \dots \mathbf{M}^6$, each one representing a combination of possible future channel states. Then we define the DTMC as $\sigma_t = \operatorname{DTMC} \left( \{ 1,\dots 6 \}, \mathbf{P}, \sigma_0 \right)$, with $\sigma_0$ being the initial Markov state and $\mathbf{P}$ being the transition matrix (e.g. learned and adjusted by machine learning techniques). This way, $\sigma_t$ points to a specific $\mathbf{M}^i$ in each time slot, so that $\mathbf{M}_t = \mathbf{M}^{\sigma_t}$.

Following \cite{PNC}, we can introduce the predictive network control (PNC) policy by defining a cost function $J$, which assigns a cost to a trajectory of control vectors $\mathbf{u}_t = (\mathbf{u}^{\intercal}_t,\mathbf{u}^{\intercal}_{t+1},..., \mathbf{u}^{\intercal}_{t+H-1})^{\intercal}$ with $H$ being the prediction horizon. We apply the usual quadratic cost function, penalizing large queue sizes as:
\begin{equation}
\label{eq:model_j}
J(\mathbf{u}_t,\mathbf{q}_t,\sigma_t) = \mathbb{E}\left[ \sum\limits^{t+H}_{i=t+1} \mathbf{q}^{\intercal}_i \mathbf{Q} \mathbf{q}_i \middle| \mathbf{q}_t,\mathbf{a}_t,\sigma_t\right]
\end{equation}

In each time slot $t$, the PNC policy minimizes $J$ under the given constraint, while explicitly considering the future development of the system. As a result, it produces an immediate control decision $u_t$. Further details can be found in \cite{PNC}.

\section{PNC Policy Implementation}
\label{sec:contribution}

The system model is inspired by \cite{PNC} and \cite{Schoeffauer002}, where the columns of $\mathbf{R}$ are modeled with discrete values, single connectivity cases are considered, and Bernoulli trials are performed to evaluate short term wireless characteristics. In this paper, we introduce:
\begin{enumerate*}[label=\itshape\alph*\upshape)]
\item a time varying $\mathbf{M}_t$ that is based on DCI, MCS, and PRBs,
\item a DC and MC connectivity model through $\mathbf{r}_{i,j}$, and 
\item a DTMC shaped by effective CQI reports following the evolution of a biased random walk.
\end{enumerate*} 
\par First, structures from the MAC layer are sent through the control plane of the MC 5G NR system to the PDCP controller. These structures shape the columns in $\mathbf{R}$, depending on $\mathcal{I}_{MCS}$ and $|\mathcal{P}|$. The entries in $\mathbf{R}$ are computed on CQI report arrivals at time $t$, based on \cite{TS38214}, as in Alg. 
\ref{alg:rt}.
\par Second, DC/MC is considered through an entry in $\mathbf{R}$ which is controlled by $\mathbf{u}^{\intercal}_t$, e.g. $\mathbf{r}_{i,2}=(-T^{\mathsf{DC}},T^{\mathsf{DC}},0,T^{\mathsf{DC}})$ for activation of links $\{\mathbf{l}_0,\mathbf{l}_1\}$ as in Fig. \ref{fig:system}. 
\par Third, $\gamma^{\text{effective}}_{k}$ indexes, $\mathcal{R}^{\text{highest}}_{k}$, are captured as wireless channel characteristics in a per sub-band approach to find $\sigma_0$ as in Alg. \ref{alg:m}. Moreover, $\gamma^{\text{effective}}_{k}$ is used to shape the weight matrices $\operatorname{diag}\{m_{j,t}\}$. For example, $m_{0,t} = 1 - P_{0,t}(T|\mathcal{I}_{MCS},\gamma^{\text{effective}}_{k})$ as in \cite[Sec. 4.3.2.1]{ieee80216m}, where $P_{0,t}$ corresponds to the error probability of a TBS at a corresponding MCS.
\newpage
\begin{algorithm}[ht]
\caption{Transport Block Size, $T_t$}\label{alg:rt}
\begin{algorithmic}[1]
\State D: Downlink control indicator (DCI)
\State $\mathcal{P}$: Set of PRBs allocated
\State $v$: Number of layers
\State $T \gets 0$
\State Read $\mathcal{I}_{MCS}$ and $Q_m$ (modulation order) from D
\Function{TBS}{$\mathcal{I}_{MCS},Q_m,|\mathcal{P}|, v$}
\State $R \gets $ LUT($\mathcal{I}_{MCS},Q_m$) in \cite[Tbl. 5.1.3.1-1]{TS38214} 
\State $N_{info} \gets |\mathcal{P}|*R*Q_m*v$
\If{$N_{info} \leq 3824$}
	\State $n=\argmax\{3, \log_2(N_{info}) - 6\}$
	\State $N_{info}^* = \argmax\{24, 2^n(\frac{N_{info}-24}{2^n})\}$
	\State $T \gets $LUT$(\mathcal{I}_{MCS},N_{info}^*)$ in \cite[Tbl. 5.1.3.2-1]{TS38214}
\Else
	\State $n \gets\log_2(N_{info} - 24) - 5$
	\State $N_{info}^* \gets \argmax\{3840, 2^n\lceil\frac{N_{info}}{2^n}\rceil\}$
	 \If{$N_{info}^* > 8424$}
		\State $C \gets \lceil \frac{N_{info}^* + 24}{8424}\rceil$		
		\State $T \gets 8C \lceil \frac{N_{info}^* +24}{8C}\rceil -24$
	\Else
		\State $T \gets 8 \lceil \frac{N_{info}^* +24}{8}\rceil -24$
	\EndIf
\EndIf
\State \textbf{return} $T$
\EndFunction
\end{algorithmic}
\end{algorithm}

\begin{algorithm}
\caption{$\sigma_0$}\label{alg:m}
\begin{algorithmic}[1]
\State $\Gamma$: Temporary map structure storing CQI reports
\State $S$: Hash map of pre-defined CQI Indexes, $ \mathcal{I}^{\mathsf{SCell}}$  
\State $\sigma_0 = 1 $: Initial state 
\State \textbf{Begin} $A6^{\mathbf{1}}_t$ Eq.\ref{eq:a61} 
\State $\Gamma \gets \mathcal{I}^{\mathsf{highest, SCell}}_k,\mathcal{I}^{\mathsf{highest, MCell}}_k,\mathcal{I}^{\mathsf{highest, SCell-neighbour}}_k$\Comment SINR mapping indexes function of Eq. \ref{eq:effectivesinr}
\Function{HashTuple}{$\Gamma$} 
\EndFunction
\Function{SearchHash}{$S,\Gamma$}
\State \textbf{return} $\mathcal{I}(S)$ 
\EndFunction
\end{algorithmic}
\end{algorithm}
\par Finally, we can define PNC as the policy for each time slot by solving the binary linear program with linear constraints \cite{PNC}, $\mathbf{u}^{*}_t$:
\begin{equation*}
\begin{aligned}
& \underset{\mathbf{u}_t}{\argmin}
&& J(\mathbf{u}_t,\mathbf{q}_t,\sigma_t) \\
&&\text{s.t.}\\
&&& \mathbf{u}_{t}\in\mathcal{U}\\
&&& \mathbf{1} \geq \left[ \mathbf{I} \otimes \mathbf{A} \right] \mathbf{u}_{t}
\end{aligned}
\end{equation*}

\section{Simulation}
\label{sec:simulation}
\par We evaluate a PDCP controller by means of SLS that follow 5G NR specifications. The simulator includes major radio resource management functionalities such as hybrid automatic repeat request and link adaptation. The network topology consists of a standard hexagonal grid of a three-sector \macrocell{} complemented with a set of outdoor \smallcells{}. \macrocells{}  and \smallcells{} are deployed at 2GHz and 3.5 GHz carrier frequency, respectively. The \macrocell{} to UE link follows the urban macro model (UMa), and the \smallcell{} to UE link follows the urban micro model (UMi). The rest of considerations are summarized in Table \ref{tbl:simulation}.
\begin{table}[h!]
\begin{center}
\caption{Simulation Summary}
\label{tbl:simulation}
\begin{tabular}{r|l}
\toprule
\textbf{Environment}        & 3GPP-Umi 3gNBs                 \\
\textbf{Traffic}            & TCP, interarrival: 10us, B=50B \\ 
\textbf{Cell Synchronicity} & Ideal\\
\textbf{Xn Latency} & 500us \\
\textbf{S1 Latency} & 10ms \\
\textbf{Buffer Size} & 20MB \\
\midrule
\textbf{MgNB Tx} & 46 dBm,  2 GHz  \\
\textbf{Channel Model} & UMa  \\
\textbf{Antenna Height} & 30m \\
\textbf{CQI Latency} & 2ms \\ 
\textbf{CQI Feedback}            & Ideal\\
\textbf{MAC Scheduler}  	& Flex TTI \\
\textbf{UE Cell Selection}  	& A3 RSRP \\
\textbf{MCS Index}  	& QPSK, 16QAM, 64QAM \\
\midrule
\textbf{SgNB Tx}            & 30 dBm, 3.5 GHz\\         
\textbf{Channel Model} & UMi  \\
\textbf{Antenna Height} & 5m \\
\textbf{CQI Latency} & 2ms  \\ 
\textbf{CQI Feedback}            & Non-Ideal\\
\textbf{MAC Scheduler}  	& Flex TTI \\
\textbf{UE Cell Selection}  	& SmallC Add A4 RSRQ \\
\textbf{MCS Index}  	& QPSK, 16QAM, 64QAM \\
\midrule
\textbf{UE Speed} & 150 km/h \\
\textbf{UE Antenna Height} & 1.6m \\
\bottomrule
\end{tabular}
\end{center}
\end{table}
\begin{figure}[th]
	\centering
	\includegraphics[width=\columnwidth]{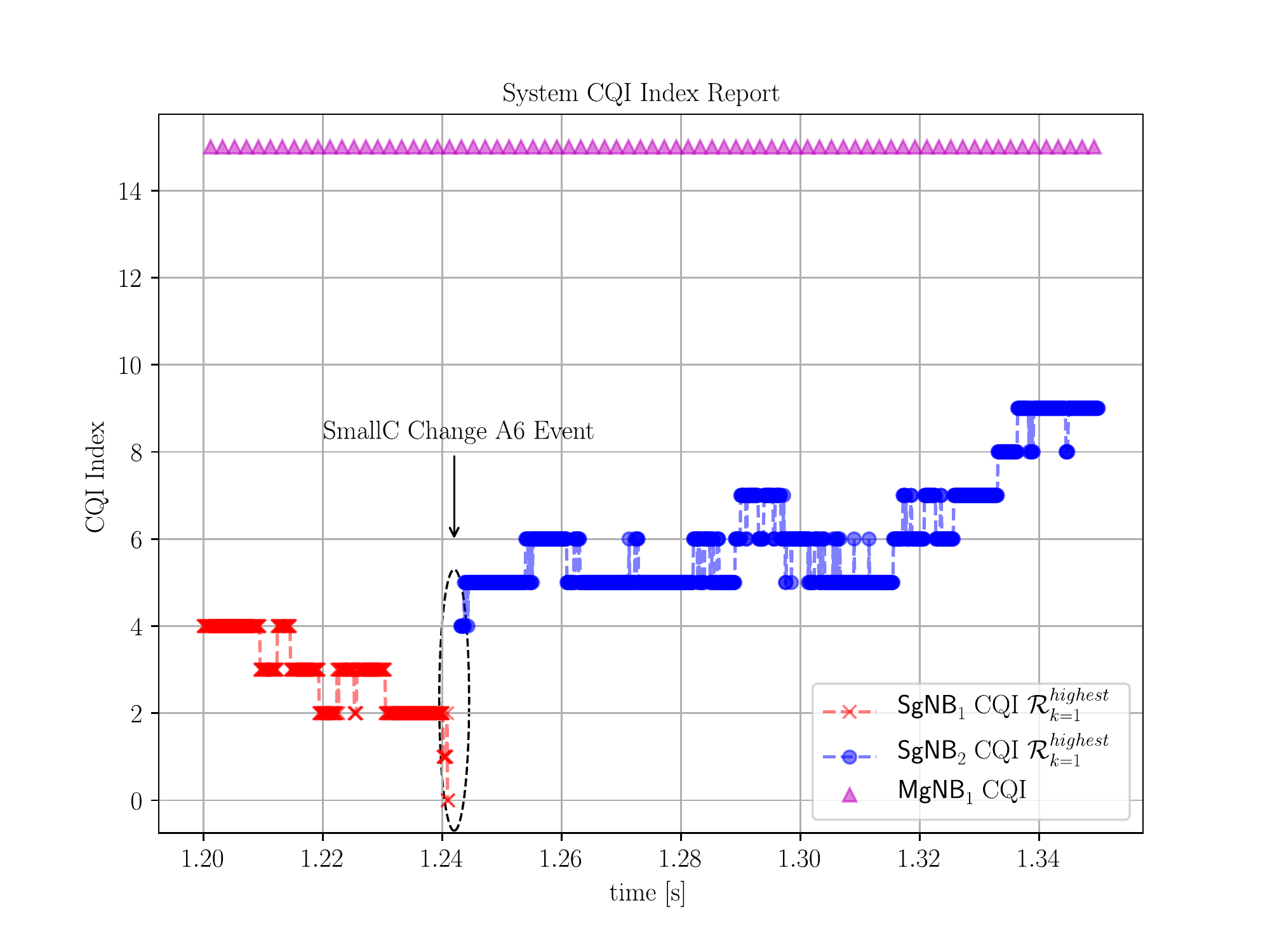}
	\caption{CQI Index}
	\label{fig:cqiperf}
\end{figure}
\par The PNC controller has knowledge of $\operatorname{diag}m_{i,t}$ in the whole system due to the Xn interfaces and CQI reports. Moreover, we assume a certain known UE trajectory at a velocity $v$ as in Fig. \ref{fig:small-cell-mobility}. The UE starts its trajectory at approximately 2 km away from the MgNB experiencing $\sim 15$ dB SINR for all its $|\mathcal{P}|$. Furthermore, it is assumed that the URLLC UE follows a high degree biased random walk. Indeed, as shown in Fig. \ref{fig:cqiperf}, the UE is moving away from $\mathsf{SgNB}_1$ and approaching $\mathsf{SgNB}_2$. 
\par Figure \ref{fig:cqiperf} shows the simulated received CQI reports from $\mathsf{MgNB}$, $\mathsf{SgNB}_1$, and $\mathsf{SgNB}_2$. It reflects the CSI report evolution in the system before and after the \smallcell{} Change event for an URLLC UE.
\par The main key performance indicator (KPI) is the E2E throughput. The E2E throughput is considered for URLLC traffic targeting a UE, and is measured from the CN to the UE PDCP layer.
\section{Performance}
\par We analyze the E2E throughput KPI gains for the system evolution. The results are obtained for an $H$ given by $A6^1_{t=1240ms}$ and $A6^2_{t=1270ms}$. The user throughput performance improves due to prediction and control of MC/DC and buffers. The user throughput is being improved due to the ability of the controller to forward PDCP PDUs to the remote, local or both MAC layer interfaces.  
\label{sec:performance}
\begin{figure}[th]
	\centering
	\includegraphics[width=\columnwidth]{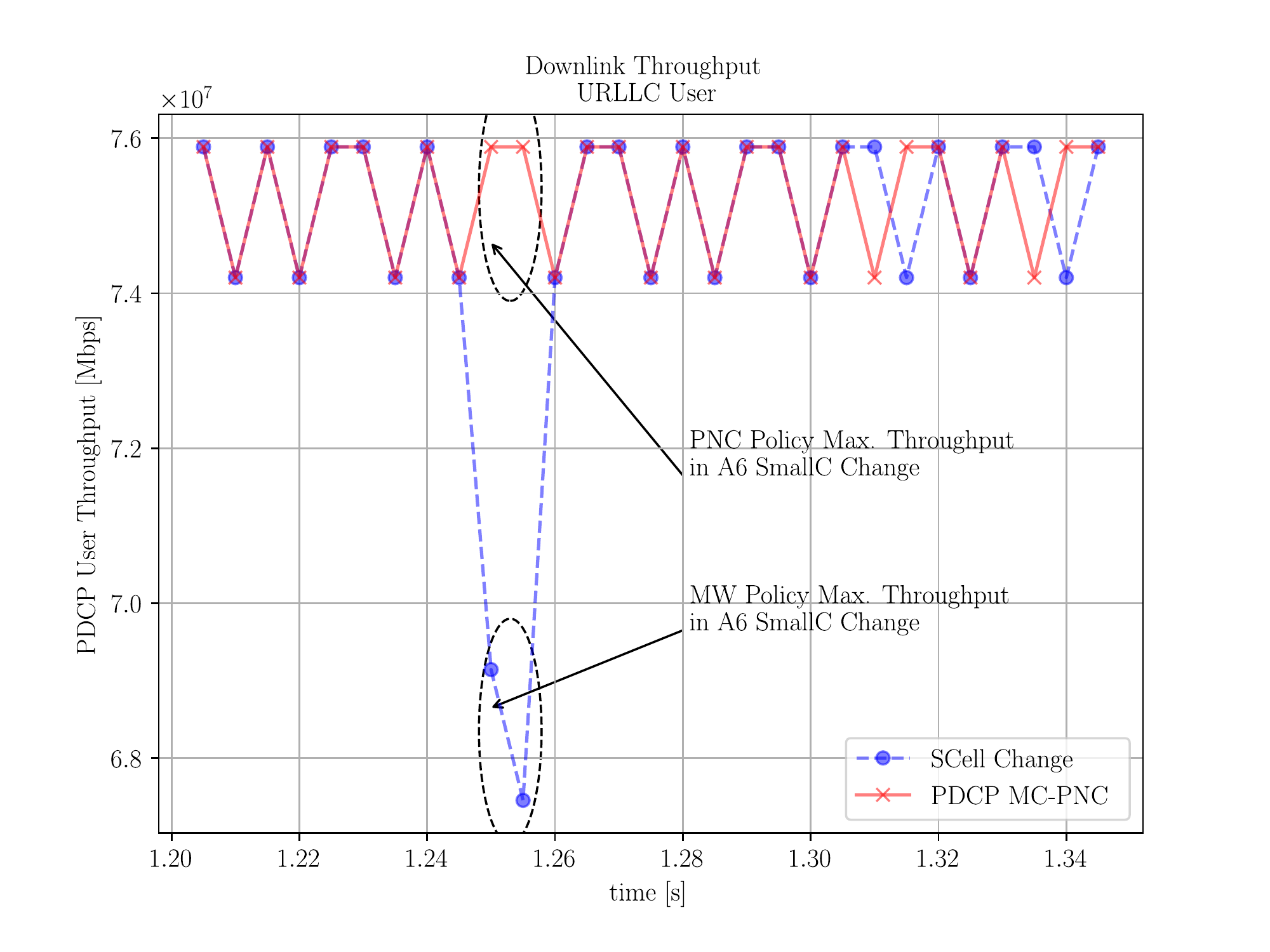}
	\caption{PNC Performance}
	\label{fig:perf}
\end{figure}
\par SLS show gains in terms of E2E throughput during the execution of a \smallcell{} Change event as shown in Fig. \ref{fig:perf}. We compare PDUs forwarding in the PDCP layer for an autonomous \smallcell{} Change based on A6 event against a PNC controlled strategy. The comparison is performed in a HetNet serving common users and URLLC users. The gains are about 10\% of E2E throughput during the execution of a \smallcell{} change process.
Figure \ref{fig:perf} shows the E2E throughput during the evolution of the system. It is depicted a maximum E2E throughput equal to $7.6$Mbps for the PNC Policy, and a maximum E2E throughput equal to $6.9$Mbps for the autonomous \smallcell{} Change.

\section{Conclusion and Outlook}
\label{sec:conclusion}

In this paper, we proposed a centralized controlled connectivity scheme for MC scenarios to enhance performance in URLLC services. The controller implements a PNC policy to forward data from the MgNB to SgNBs or UEs. The proposed scheme keeps track of wireless channel variations and buffer status. The wireless channel variations through means of CQI reports, and the buffer status through Xn application protocol reports.    

We addressed concepts of multi-connectivity, mobility and channel state information in the 5G NR, to further highlight the potential gain offered by a PNC policy implementation applied to a PDCP forwarding layer. The gains appear as an optimized throughput during  \smallcell{} change.       
Further work will focus on spectral efficiency analysis over the different control selections applied to the PDCP forwarding layer, as well as signalling overhead analysis in the Xn interface. A future model will consider an interference mitigation receiver \cite{Ohwatari} to map CQI reports.

\section{Acknowledgements}
\label{sec:ack}

This work was performed in the framework of the H2020 project ONE-5G (ICT-760809) receiving funds from the EU. The authors would like to acknowledge the contributions of their colleagues in the project, although the views expressed in this work are those of the authors and do not necessarily represent the project.

In addition, part of this work is financially supported by the German Research Foundation (DFG) within the priority program SPP1914.
\bibliographystyle{IEEEtran}
\bibliography{biblio}

\end{document}